\documentclass[aps,preprint,superscriptaddress,amsmath,groupedaddress,showpacs,tightenlines,nofootinbib]{revtex4-1}

\usepackage{graphicx}  %
\usepackage{amsmath}
\usepackage{amssymb}
\usepackage{bm}  %
\usepackage{tikz}
\usetikzlibrary{calc,patterns,angles,quotes}

\usepackage{feynmp}
\usepackage{epsfig}
\usepackage{times}
\usepackage{color}
\usepackage{hyperref}
\usepackage[section]{placeins}

\newcommand{\bea}{\begin{eqnarray}}
\newcommand{\eea}{\end{eqnarray}}
\newcommand{\beq}{\begin{equation}}
\newcommand{\eeq}{\end{equation}}
\newcommand{\bqa}{\begin{eqnarray}}
\newcommand{\eqa}{\end{eqnarray}}

\def\mqo2{{\!\!\!}}

\renewcommand{\Re}{{\rm Re\,}}

\newcommand{\expval}[1]{\langle #1 \rangle}
\newcommand{\matrixel}[3]{\left< #1 \mvph{#2#3} \right| #2 \left| #3 \mvph{#1#2} \right>} 
\newcommand{\alignmentpoint}{&}
\newcommand{\mvph}[1]{\vphantom{\def\alignmentpoint{}#1}}

\def\fun#1#2{\lower3.6pt\vbox{\baselineskip0pt\lineskip.9pt
  \ialign{$\mathsurround=0pt#1\hfil##\hfil$\crcr#2\crcr\sim\crcr}}}

\newcommand{\ifcitation}[2]{#2}
\newcommand*{\prv}{Phys. Rev.} 
\newcommand*{\jpg}{\ifcitation{JPG}{J. Phys. G}} 
\newcommand*{\jpb}{\ifcitation{JPG}{J. Phys. B}} 

\newcommand*{\physrep}{Phys. Rep.} 
\newcommand*{\plb}{Phys. Lett. B} 
\newcommand*{\arnps}{\ifcitation{ARNPS}{Ann. Rev. Nucl. Part. Sci.}}

\newcommand*{\epja}{\ifcitation{EPJA}{Eur. Phys. J. A}}
\newcommand*{\fbs}{Few Body Syst.}
\newcommand*{\ppnp}{Prog. Part. Nucl. Phys.}
\newcommand*{\ptp}{Prog. Theor. Phys.}
\newcommand*{\rpp}{Rep. Prog. Phys.}

\newcommand*{\njp}{\ifcitation{NJP}{New J. Phys.}}
\newcommand*{\npa}{Nucl. Phys. A}
\newcommand*{\np}{\ifcitation{NP}{Nat. Phys.}}
\newcommand*{\cphc}{\ifcitation{CPC}{ChemPhysChem}}

\definecolor{dimerblue}{HTML}{1733ed}
\definecolor{trimerred}{HTML}{ed171b}
\definecolor{tetramergreen}{HTML}{339530}

\begin{document}

\title{Efimov Universality with Coulomb Interaction}
\author{C. H. Schmickler}
\email{schmickler@theorie.ikp.physik.tu-darmstadt.de}
\affiliation{Institut f\"ur Kernphysik, Technische Universit\"at Darmstadt,
64289\ Darmstadt, Germany}
\affiliation{Nishina Center, RIKEN, Saitama 351-0198, Japan}
\author{H.-W. Hammer}
\affiliation{Institut f\"ur Kernphysik, Technische Universit\"at Darmstadt,
64289\ Darmstadt, Germany}
\affiliation{ExtreMe Matter Institute EMMI, GSI Helmholtzzentrum f\"ur 
Schwerionenforschung, 64291\ Darmstadt, Germany}
\author{E. Hiyama}
\affiliation{Department of Physics, Kyushu University, Fukuoka 819-0395, Japan}
\affiliation{Nishina Center, RIKEN, Saitama 351-0198, Japan}
\date{\today}

\begin{abstract}
  The universal properties of charged particles are modified
  by the presence of a long-range Coulomb interaction.
  We investigate the modification of Efimov universality
  as a function of the Coulomb strength using the Gaussian
  Expansion Method. The resonant short-range interaction
  is described by Gaussian potentials to which a Coulomb potential is added.
  We calculate binding energies and root mean square radii 
  for the three- and four-body systems of charged particles and present
  our results in a generalised Efimov plot. 
  We find that universal features can still
  be discerned for weak Coulomb interaction,
  but break down for strong Coulomb interaction.
  The maximum root-mean-square radius of the system
    decreases as the strength of the Coulomb interaction is
    increased  and the
  probability distributions of the states become more concentrated inside
  the Coulomb barrier. As an example,
  we apply our universal model to nuclei with an $\alpha$ cluster substructure. 
  Our results point to strong non-universal contributions in that sector. 
\end{abstract}

\smallskip
\maketitle

\section{Introduction}
\label{sec:intro}

Few-body systems of strongly-interacting particles 
  may show universal properties
independent of the details of their interaction at short
distances~\cite{Efimov:1970zz,Braaten:2004rn}.
The simplest example is given by identical bosons with large
S-wave scattering length $a$ and mass $m$.
If $a$ is positive and much larger than the range of
the interaction $r_0$,  the system has a shallow
two-body bound state with binding energy
\beq
B_2 = \frac{\hbar^2}{m a^2}(1+{\cal O}(r_0/a))\,,
\label{eq:b2}
\eeq
and mean-square separation
$a^2/2$.

\begin{figure}[htb]
\centering
\includegraphics[width=6cm,clip=true]{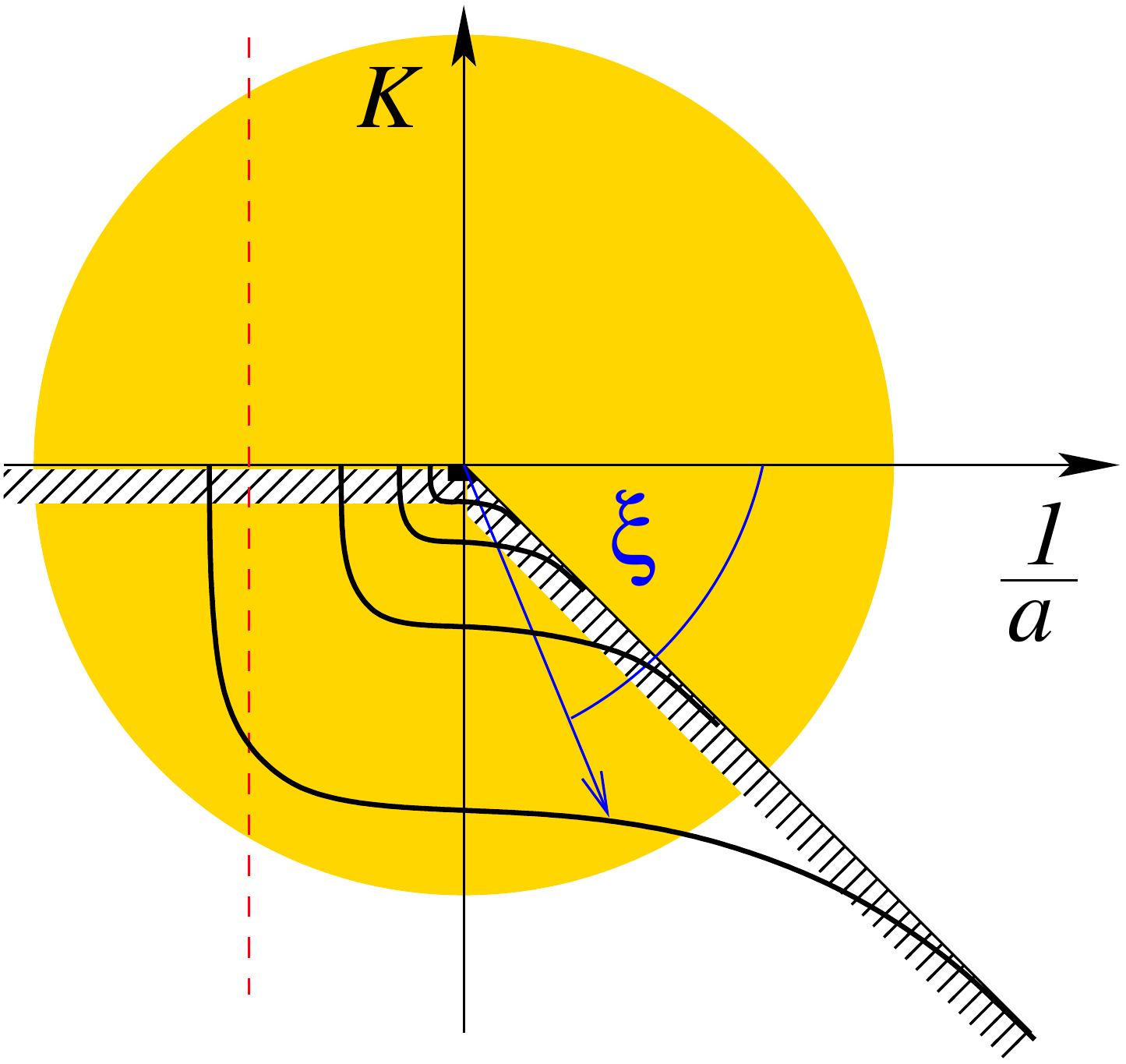}

\caption{Efimov plot of the three-body bound state spectrum. The
energy variable $K={\rm sgn}(E)\sqrt{m|E|}$ is shown as a
function of the inverse scattering length $1/a$. The solid lines
indicate the Efimov states while the hashed areas give the scattering
thresholds. The dashed vertical line illustrates a 
specimen of a physical system with fixed scattering length.
}
\label{fig:efiplot}
\end{figure}

In systems of three and more bosons, a three-body parameter
$\kappa_*$ is required to characterize the system.
For fixed scattering length, this implies universal correlations between
different few-body observables such as the Phillips \cite{Phillips:1968zze}
and Tjon lines \cite{Tjon:1975sme} if the two-body interaction has the
same on-shell properties. In the three-body system, the
Efimov effect \cite{Efimov:1970zz} generates a universal spectrum
of three-body bound states with binding energy
\beq
B_3 (1+{\cal O}(r_0/a)) = - \frac{\hbar^2}{m a^2} +
\left[e^{-2 \pi n} f(\xi)\right]^{1/s_0}
\,\frac{\kappa_*^2}{m}\,,
\label{B3-Efimov}
\eeq
where $s_0 = 1.00624...$ is a transcendental number,
the angle $\xi$ is defined by
$\tan \xi = - (mB_3)^{1/2} \, a /\hbar \,$,
and $f(\xi)$ is a universal function with   $f ( -\pi/2 )=1$
(see Ref.~\cite{Braaten:2002sr} for more details).
This spectrum is represented in the \textit{Efimov plot}
shown in Fig.~\ref{fig:efiplot}.
It consists of the two-dimensional plane spanned by the energy variable
$K={\rm sgn}(E)\sqrt{m|E|}$ and the inverse scattering length $1/a$.
The three-body parameter $\kappa_*$ is not shown explicitly, as
it only sets the overall scale of the plot. The shaded area of
radius $1/r_0$ shows the region where universality is expected to hold,
while the solid lines indicate the Efimov states and  the hashed areas give
the scattering thresholds. The dashed vertical line illustrates
 a  specimen of a
system with fixed scattering length. The spectrum displays
invariance under discrete scaling transformations with the scaling
factor $\exp{(\pi/s_0)}\approx 22.7$: the knowledge of
the scattering length dependence of one state determines the scattering
length dependence of all other states.

In the unitary limit $1/a=r_0=0$, the six-dimensional
three-body problem reduces to a
one-dimensional Schr\"odinger equation in the hyperradius\footnote{
  The hyperradius $\rho_{\text{hyp}}$ is the mean separation of the three
  particles, i.e. $\rho_{\text{hyp}}^2=(r_{12}^2+r_{13}^2+r_{23}^2)/3$ for
  equal masses.}
with an attractive inverse square potential.
The Efimov spectrum then becomes geometric:
\beq
B_3 = \left(e^{-2 \pi/ s_0} \right)^{n-n_*} \frac{\kappa_*^2}{m}\,,
\label{B3-Efimov-uni}
\eeq
where $\kappa_*$ is identified as
the binding momentum of the state with label $n_*$,
and the binding momenta of
neighbouring states differ by the scaling factor $\exp{(\pi/s_0)}\approx 22.7$. 
Two universal four-body states with binding energies
\beq
\label{eq:scale-tet}
B_4^{0} = 4.610(1)\,B_3 \quad \mbox{ and }\quad
B_4^{1} = 1.00227(1)\,B_3 
\eeq
are attached to each Efimov
state~\cite{Platter:2004he,Hammer:2006ct,vStecher:2008,deltuva2010,deltuva2012hpw}. This pattern is expected to hold for higher-body states
as well~\cite{vonStecher:2011zz,Gattobigio:2012tk,Kievsky:2014dua,Bazak:2016wxm}.
However, at some point their size becomes so small that they leave the universal
region. The properties of these systems are generically 
referred to as \textit{Efimov physics}~\cite{Braaten:2004rn,Naidon:2016dpf}.
In ultracold atoms, signatures of these universal states
have been observed for up to five particles
\cite{Ferlaino:2010viw,Zenesini:2013zz}.

The Efimov effect is also relevant for the
binding of weakly-bound states in nuclear and particle
physics, such as hadronic molecules and halo nuclei.
For reviews of the status of Efimov physics in these systems,
see Refs.~\cite{Jensen:2004zz,Hammer:2010kp,Hammer:2017tjm}.
In systems with two or more charged constituents, however,
the short-range strong interaction is accompanied by long-range
Coulomb interactions. The effect of Coulomb
interactions on Efimov physics and discrete
scale invariance is therefore an important issue.
Qualitatively, one expects shallow states with sizes
of the order of the Bohr radius or larger to be Coulombic
while deeper states with size smaller than the Bohr radius
are Efimovian in character \cite{Efimov:1990rzd}.
In the simpler case of a two-body system with
inverse square interaction, these expectations 
were confirmed by explicit calculations
\cite{Barford:2002je,Hammer:2008ra}. An analysis of the
gross properties of three-body halo nuclei with Coulomb
interaction was carried out in Ref.~\cite{Fedorov:1994}.
Moreover, various studies have focused
on the treatment of Coulomb interactions in short-range effective
field theories of the three-nucleon system (see, e.g.,
Refs.~\cite{Ando:2010wq,Konig:2011yq,Vanasse:2014kxa,Konig:2015aka}).
In particular, K\"onig et al.~\cite{Konig:2016utl} have considered
the universal properties of three- and four-nucleon systems
in a strict expansion around the unitary limit, where Coulomb
forces are subleading. Their counting is based on the assumption 
that light, and possibly heavier, nuclei are bound weakly enough
to be insensitive to the details of the interactions but strongly
enough to be insensitive to the exact size of the two-nucleon system. 
By construction, their counting is not applicable at very low energies
where the finite scattering lengths strongly influence the spectrum
and the Coulomb interaction cannot be treated in perturbation theory.

Here, we attempt to fill a gap in previous calculations and focus explicitly
on the modifcation of Efimov universality and discrete scale
invariance in the presence of Coulomb forces. We follow the standard
Efimov scenario and include the scattering lengths at leading order.
Our results are summarized in generalised Efimov plots.
The paper is organised as follows.
First we will introduce the method that is used in our calculations, 
the Gaussian Expansion Method, in Sec.~\ref{sec:gem}, our Hamiltonian in Sec.~\ref{sec:interaction},
and the Coulomb-modified scattering length
in Sec.~\ref{sec:coulombmodifiedsl}. 
We will then show our results in natural units in 
Sec.~\ref{sec:spectrum} and Sec.~\ref{sec:structure}. 
In the last part (Sec.~\ref{sec:applications}), we apply our calculations to the $N\alpha$ system and 
discuss the difficulties arising there. We then summarize our findings 
and indicate possible future projects in the outlook (Sec.~\ref{sec:outlook}).

\section{Method}

\subsection{Gaussian Expansion Method}
\label{sec:gem}

We use the Gaussian Expansion Method (GEM) \cite{hiyamakino2003} to calculate 
binding energies and wave functions of the systems of $N$ identical charged bosons
under investigation.

Successfull applications of this method include various hypernuclear systems
up to $N=5$ \cite{hiyamakamimura1996,hiyamakamimura1997,hiyamakamimura1999,hiyamakamimura2002,
hiyamakamimura2002b,hiyamakamimura2010} and systems of atomic $^4$He \cite{hiyamakamimura2012,hiyamakamimura2012b,
hiyamakamimura2014}. 

\begin{figure}[tb]
\bigskip
\includegraphics{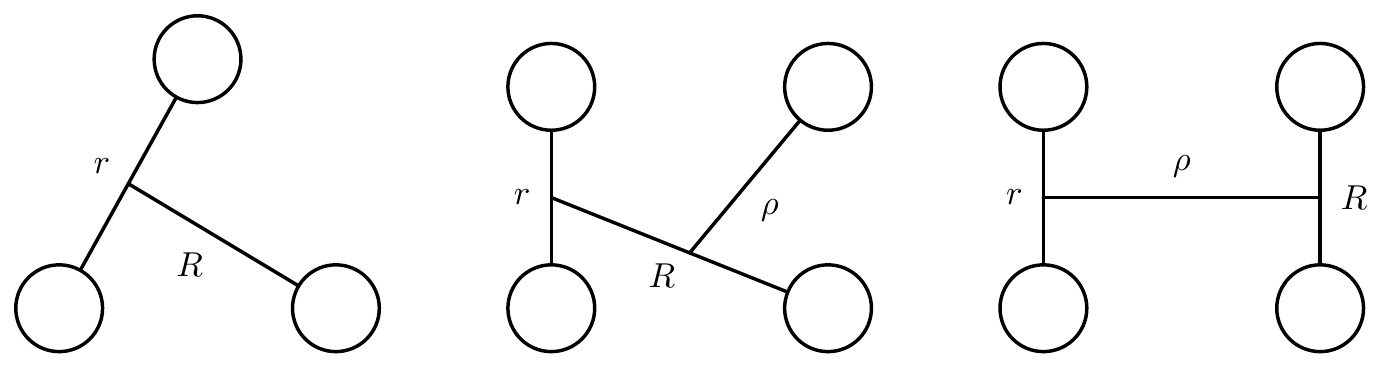}
\caption{Illustration of Jacobi coordinate sets used in our calculations.}
 \label{jacobicoordinates}
 \end{figure}

The GEM utilises products of Gaussian basis functions to describe the system in 
terms of Jacobi coordinates. The Jacobi coordinates are illustrated in 
Fig.~\ref{jacobicoordinates}. There is only one configuration (``channel'') in the three-body 
system and two in the four-body system, because we treat identical particles. 

We then solve the Schr\"odinger equation
\begin{equation}
 (H-E)\Psi = 0.
\end{equation}
The details of the Hamiltonian $H$ can be found in Sec.~\ref{sec:interaction}. 
The total wave function $\Psi$ is comprised of Gaussian functions as 
explained in the following.

The basis functions for each Jacobi coordinate take the form of
\begin{align}
\phi_{nlm}({\bf r}) & =  \phi^{\rm G}_{nl}(r)\:Y_{lm}({\widehat {\bf r}}) ,&
 \phi^{\rm G}_{nl}(r) &=  N_{nl}\,r^l\:e^{-(r/r_n)^2},
\label{eq:3gaussa}\\
\psi_{NLM}({\bf R}) & =  \psi^{\rm G}_{NL}(R)\:Y_{LM}({\widehat {\bf R}}) , &
 \psi^{\rm G}_{NL}(R) & = 
N_{NL}\,R^L\:e^{-(R/R_N)^2},
\label{eq:3gauss} \\
\xi_{\nu \lambda \mu}(\mbox{\boldmath $\rho$}) & =  
 \xi^{\rm G}_{\nu \lambda \mu}(\rho)\:Y_{\lambda \mu}
({\widehat {\mbox{\boldmath $\rho$}}}) , &
 \xi^{\rm G}_{\nu \lambda}(\rho) & = 
N_{\nu \lambda}\,\rho^\lambda\:e^{-(\rho/\rho_\nu)^2},
\label{eq:3gaussb}
\end{align}
where
$N_{nl}$, $N_{NL}$, $N_{\nu \lambda}$  denote the normalization constants.
The range parameters of the Gaussians are chosen to form 
a geometric progression:
\begin{align}
 r_n& =r_{\rm min}\, a^{n-1},&                   a& =\left(\frac{r_{\mathrm{max}}}{r_{\mathrm{min}}}      \right)^{\frac{1}{n_{\mathrm{max}}-1}}& (n& =1,...,n_{\rm max})\;, \\
 R_n& =R_{\rm min}\, A^{N-1},&                   A& =\left(\frac{R_{\mathrm{max}}}{R_{\mathrm{min}}}      \right)^{\frac{1}{N_{\mathrm{max}}-1}}& (N& =1,...,N_{\rm max})\;, \\
 \rho_n& =\rho_{\rm min}\, \alpha^{\nu-1},& \alpha& =\left(\frac{\rho_{\mathrm{max}}}{\rho_{\mathrm{min}}}\right)^{\frac{1}{\nu_{\mathrm{max}}-1}}& (\nu& =1,...,\nu_{\rm max})\;.
\label{eq:progR}
\end{align}

This ensures that short-range correlations and long-range contributions can 
both be described very well and makes it especially well-suited to the problem 
discussed in this study.

  The values for $r_\text{min}, R_\text{min}, \rho_{min}$
  and $r_\text{max}, R_\text{max}, \rho_{max}$ etc. are of the order of
  $0.1r_0$ and $1000r_0$, respectively. The number of basis functions
  $n_\text{max}, N_\text{max}, \nu_\text{max}$ etc. are around 
  25 for most calculations.
This provides an accuracy on the level that can be discerned on 
the plots.

The total wave function is then 
\begin{equation} \label{wave}
 \begin{split}
  \Psi^{\rm trimer}_{JM} & =  \sum_{c} \sum_{n_c,N_c}\sum_{\ell_c,L_c} C^{(c)}_{n_c \ell_c N_c L_c} \sum_{b\in p(c)} 
  [\phi^{(c)}_{n_c \ell_c}({\bf r}_b) \psi^{(c)}_{N_c L_c}({\bf R}_b)]_{JM} \\
\Psi^{\rm tetramer}_{JM} & =  \sum_{c} \sum_{n_c,N_c,\nu_c}\sum_{\ell_c,L_c,\lambda_c}  
C^{(c)}_{n_c \ell_c  N_c L_c \nu_c \lambda_c } \sum_{b\in p(c)} [[\phi^{(c)}_{n_c \ell_c}({\bf r}_b) \psi^{(c)}_{N_c L_c}({\bf R}_b)]_I \xi_{\nu_c \lambda_c}^{(c)}
(\mbox{\boldmath $\rho$}_b)]_{JM},
 \end{split}
\end{equation}
where $c$ is the configuration channel as illustrated in Fig.~\ref{jacobicoordinates} and 
$p(c)$ are all possible permutations of particles in a given channel. 
$C^{(c)}_{n_c\ell_cN_cL_c}$ and $C^{(c)}_{n_c\ell_cN_cL_c\nu_c\lambda_c}$  are coefficients. 

Since we are interested in the leading order results we take all angular momentum quantum 
numbers ($\ell, L, \lambda$) to be zero. This also takes care of the symmetrization 
of our identical bosons. We are also only calculating states with total angular momentum zero.

Note that even though we take all angular momentum quantum numbers to be zero, the 
symmetrization of the wave function, i.e. adding wave functions with different pairing 
for the Jacobi coordinates, leads to an implicit incorporation of some higher-partial wave 
contributions. 

We tested our calculations against some results of a correlated Gaussian method (SVM) that 
explicitly takes into account higher partial waves, and found very good agreement. (See for 
more detail the supplementary material of \cite{schmicklerhammervolosniev2019}.) This suggests 
to us that the important part of the higher partial wave contributions is already contained 
in our calculations. 

Furthermore, we also calculated with (explicit) higher partial waves for a few select points and 
found a difference to the result without higher partial waves that was smaller than 
what could be discerned on the plots we show. This gives us further confidence in our assessment.

With this the Schr\"odinger equation for the three-body system is 
\begin{equation}
\begin{split}
 \sum_{c}\sum_{n_cl_cN_cL_c} &\sum_{b\in p(c)}\sum_{b'\in p(c')}C^{(c)}_{n_cl_cN_cL_c} \\ &\times 
 \matrixel{[\phi^{(c')}_{n_{c'} \ell_{c'}}({\bf r}_{b'}) \psi^{(c')}_{N_{c'} L_{c'}}({\bf R}_{b'})]_{JM}}{(H - 
 E)}{%
 [\phi^{(c)}_{n_c \ell_c}({\bf r}_b) \psi^{(c)}_{N_c L_c}({\bf R}_b)]_{JM}
} = 0,
\end{split}
\end{equation}
which due to the non-orthogonality of the basis states corresponds to a general 
eigenvalue problem:
\begin{equation}
 \sum_{\tilde n}(H_{\tilde n'\tilde n} - EN_{\tilde n'\tilde n})C_{\tilde n} = 0,
\end{equation}
where $N_{\tilde n'\tilde n'}$ is a normalization matrix and $\tilde n$ comprises 
all indices that are summed over. The matrix elements can be calculated analytically 
for all potentials used here and we solve the equation using standard linear 
algebra methods. The four-body system can be treated analogously.

\subsection{Interaction}
\label{sec:interaction}
In this subsection, we specify our interaction Hamiltonian.
We consider equal mass particles of mass $m$.
The two-body Hamiltonian contains a Gaussian potential to model the
short-range interaction and a long-range Coulomb potential,
\begin{equation}
 H_2 = -\frac{\hbar^2}{2\mu}\frac{\partial^2}{\partial r^2} +  V_0 e^{-\frac{r^2}{2r_0^2}} + \hbar cZ_1Z_2\frac{\alpha}{r}\, ,
\end{equation}
where $\mu=m/2$ is the reduced mass while $V_0$ and $r_0$ determine the strength
and the range of the Gaussian potential, respectively.
The short-range part has been used in many works to 
investigate the Efimov effect \cite{gattobigiokievsky11,blumeyan2014,schmicklerhammer2017}. It is well 
suited to the investigation of universal physics because it can be regarded
as a contact term with Gaussian smearing. If one considers
only energies that are small compared to the natural scale
$E_s = \hbar^2/(2\mu r_0^2)$,
it corresponds to the leading order of an effective
theory for large scattering length. As we are only interested 
in the universal behavior and not in the short-range details of any
particular interaction, we will use this Hamiltonian in our analysis. 
Adding the Coulomb interaction to this Hamiltonian is straightforward.
$Z_1$ and $Z_2$ are the charge numbers of particles 1 and 2, respectively,
and $\alpha\approx 1/137$ is the fine structure constant. 

The generalisation of the Hamiltonian to systems of three and four particles
is also straightforward:
\begin{equation}
 H_3 = -\frac{\hbar^2}{2\mu_r}\nabla_r^2 - \frac{\hbar^2}{2\mu_R}\nabla_R^2 
  +V_0\sum^3_{i<j}e^{-\frac{r_{ij}^2}{2r_0^2}} + \hbar c \sum^3_{i<j}Z_iZ_j\frac{\alpha}{r_{ij}}\,,
\end{equation}
and
\begin{equation}
 H_4 = -\frac{\hbar^2}{2\mu_r}\nabla_r^2 - \frac{\hbar^2}{2\mu_R}\nabla_R^2 
  - \frac{\hbar^2}{2\mu_\rho}\nabla_\rho^2
  +V_0\sum^4_{i<j}e^{-\frac{r_{ij}^2}{2r_0^2}} + \hbar c \sum^4_{i<j}Z_iZ_j\frac{\alpha}{r_{ij}}\,,
\end{equation}
with $r_{ij}$ the distance between particle $i$ and $j$.
In the kinetic part, we have used the Jacobi coordinates $r, R, \rho$
with the corresponding reduced masses $\mu_r$, $\mu_R$, and $\mu_\rho$.

It is useful to express our Hamiltonian in natural units by dividing out
the natural energy scale $E_s$. Using the reduced distance
$\tilde r = r/r_0$, this gives us for the two-body Hamiltonian
\begin{equation}
 \tilde H_2 = -\nabla^2_{\tilde r} +  \tilde V_0 e^{-\frac{\tilde r^2}{2}} + \frac{\tilde c}{\tilde r}\,,
 \label{eq:universalH}
\end{equation}
with the definitions $\tilde V_0 = V_0/E_s$ and $\tilde c = c_c/c_s$ with
$c_c = \hbar c Z_i Z_j \alpha$ and $c_s = \hbar^2/(mr_0)$.
In analogy to $E_s$, $c_s$
can be regarded as natural scale of the Coulomb coupling strength. 

It is now obvious that the same result for different $r_0$ can be obtained, as long as 
$\tilde c$ stays the same.
Note, however, that expressing the charge in terms of $\tilde c$: 
$Z_iZ_j = \tilde{c} \hbar/(c\alpha mr_0)$.
Thus the same value of $\tilde c$ corresponds to different
physical charges for different $r_0$ and a value for $r_0$
has to be chosen when comparing to physical systems. We will 
discuss this issue in more details in Sec.~\ref{sec:applications}
and present our approach to choosing $r_0$. First, however, we 
will concentrate on observables in natural units.
We will present results for fixed values of $\tilde c$, which 
can be viewed as representatives of different universality classes.

For completeness, we note that in Sec.~\ref{sec:applications} 
we also include a short-range three-body force in order to fit the states
to real physical systems. It has the form
\begin{equation}
 V_3 = W_0 \sum_{i\neq j \neq k \neq i}^{N}e^{-\frac{r_{ij}^2 + r_{jk}^2 + r_{ki}^2}{16r_0^2}}.
 \label{eq:threebodyforce}
\end{equation}

\subsection{Coulomb-modified Scattering Length}
\label{sec:coulombmodifiedsl}

The scattering length is an important parameter in the
discussion of the Efimov effect. 
However, with the long-range Coulomb interaction present, the ``usual''
scattering length $a$ does not exist and one has to define a 
Coulomb-modified scattering length $a_C$.
It is related to the phase shift $\tilde\delta_0(p)$ between ingoing 
and outgoing Coulomb waves, much like the scattering length $a$ is 
related to the phase shift between ingoing and outgoing 
free waves. 

The Coulomb-modified effective range expansion for
angular momentum $\ell=0$ is 
\begin{equation}
 C^2_{\eta,0}p\cot\tilde\delta_0(p) + \gamma h(\eta) = -\frac{1}{a_C} + \frac{1}{2}r_Cp^2 + \ldots, 
\end{equation}
with $C^2_{\eta,0} = 2\pi\eta/(e^{2\pi\eta}-1)$ , $h(\eta) = \Re
[\Gamma'(i\eta)/\Gamma(i\eta)]-\log(\eta)$, $\gamma = 2\frac{\mu c^2}{\hbar c}\alpha Z_1Z_2$,
and $\eta = \gamma/(2p)$. Here, $p$ and $\gamma$ are wave numbers and have 
the unit $\text{fm}^{-1}$. More details can be found in
Refs.~\cite{koenigphd,bethe1949,vanhaeringenkok1982}. 

This definition introduces some complications in practical calculations.
For example, it is not 
possible to numerically calculate
the Coulomb-modified scattering length $a_C$ directly at 
zero energy. This means that we have to extrapolate to zero to obtain the 
scattering length. The extrapolation, however, is much more accurate than the 
accuracy of the results of our main calculation, so uncertainties stemming from 
this are negligible.

 Note that the relation between the dimer energy
  and $a_C$ is, in general, not given by replacing $a$ with $a_C$
  in Eq.~(\ref{eq:b2}) since the dimer energy depends on both $a_C$ and
  $\gamma$. (See Ref.~\cite{schmicklerhammervolosniev2019} for explicit
  expressions in some limiting cases.)
  Here, we calculate the dimer energy numerically by solving the
  Schr\"odinger equation.

\section{Generalised Efimov Plot}
\label{sec:spectrum}
\begin{figure}[tb]
\bigskip
\includegraphics{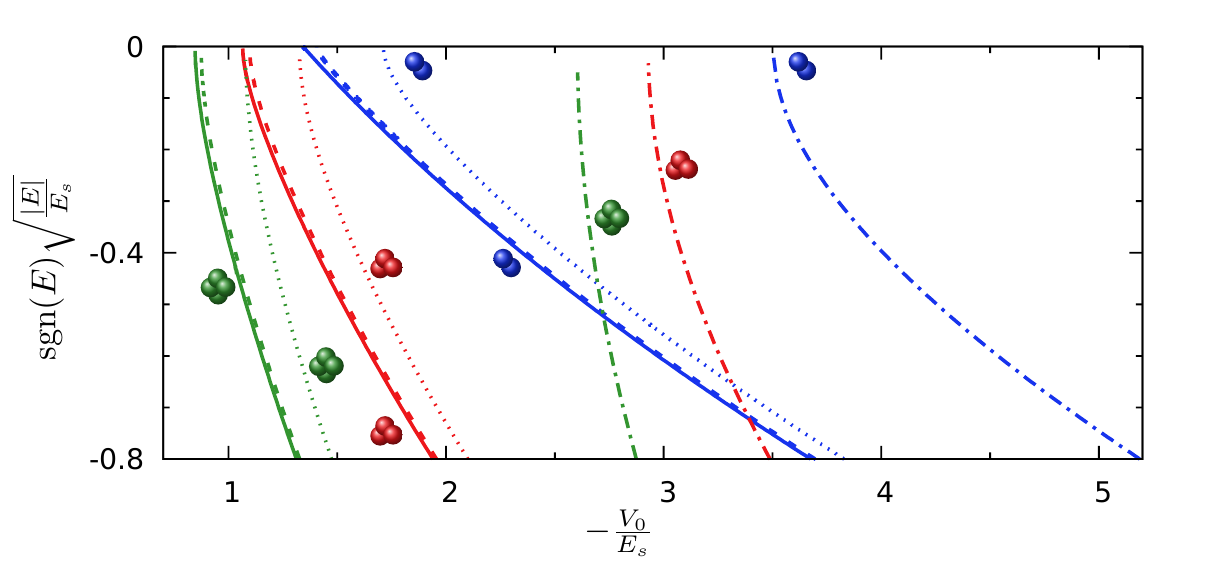}
\caption{Comparison of the generalised Efimov plot for different Coulomb interaction strength parameters in terms of $V_0$. 
The solid lines are without Coulomb interaction, the dashed lines for $\tilde c=0.007$, the dotted 
lines for $\tilde c=0.07$ and the dash-dotted lines for $\tilde c = 0.7$.
The blue lines (always the right-most of a given set) are the dimers, the red lines (always in the middle for a given set) the trimers
and the green lines (always the left-most of a given set) the tetramers. Excited states 
are  not shown to avoid confusion. }
\label{v0plot}
\end{figure}

We start by summarizing our results for different potential strengths $V_0$ in
the generalised Efimov plot of Fig.~\ref{v0plot}.
Plotted in natural units, the values for different ranges 
of the Gaussian potential $r_0$ lie on the same line, which is clear from 
Eq.~(\ref{eq:universalH}). The Efimov plot is almost unchanged
for very small values of the Coulomb interaction strength parameter $\tilde c$.
Figure~\ref{v0plot} also shows that states do become shallower with
increasing repulsive Coulomb interaction as expected.
In the following, we will show the energies as a function of the
inverse Coulomb-modified scattering length $1/a_C$, which is
an observable quantity and thus better suited to discuss universal
properties.

  In the following, we show results for $\tilde c = 0.007, 0.07, 0.7$, which
  correspond, e.g., to particles with the $\alpha$ particle mass and a charge
  of $0.2, 0.7$ and $2.3$, respectively, if $r_0$ is assumed to
  be $1\,\text{fm}$. If we assume $r_0 = 2\,\text{fm}$, 
  the corresponding charges are  $0.16, 0.5, 1.6$. 

  These values were chosen to achieve an even spread over the region that was
  accessible to our calculations and to showcase the qualitatively different
  results encountered within this range.

\begin{figure}[tb]
\bigskip
\includegraphics{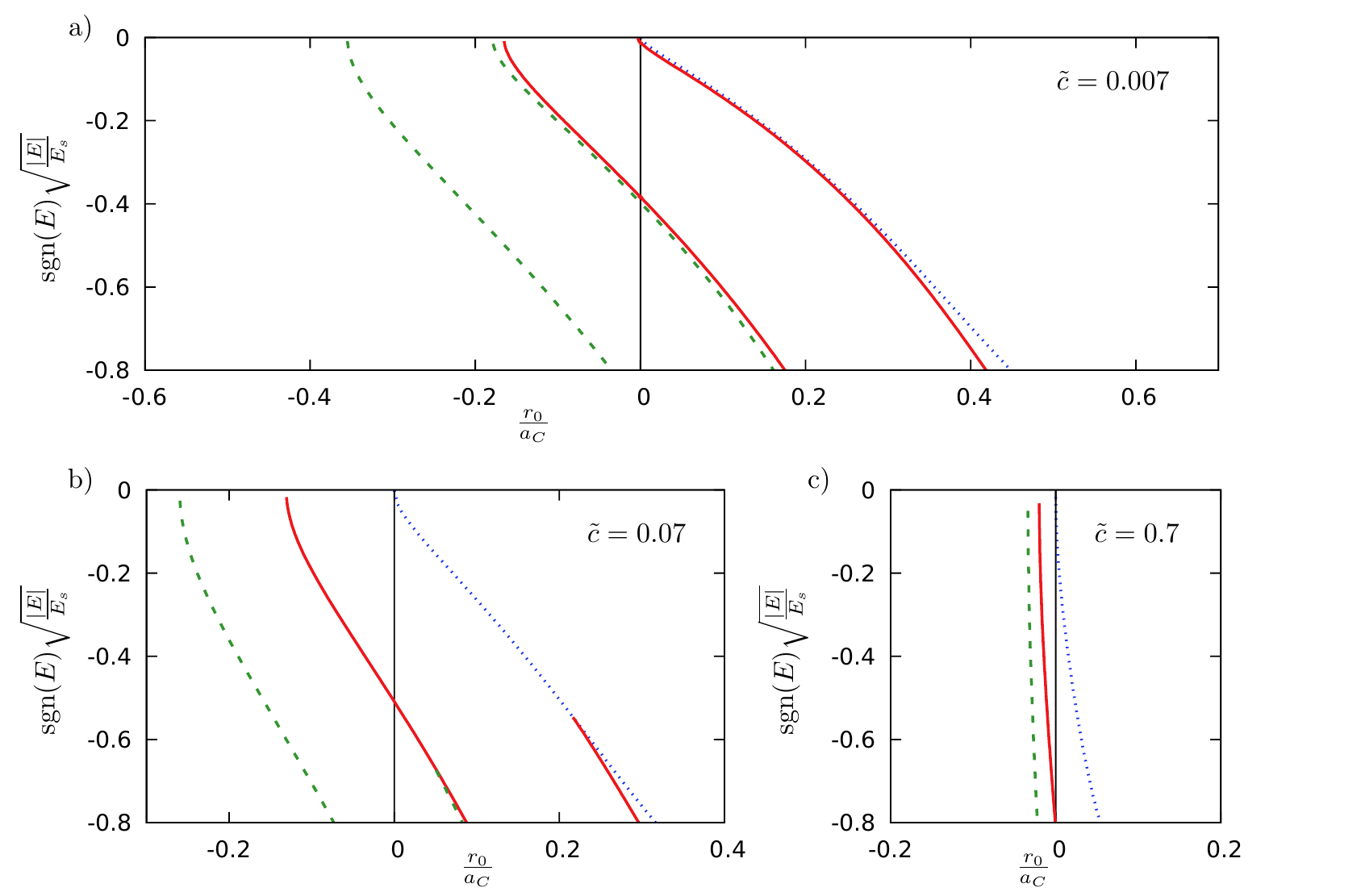}
\caption{Generalised Efimov plot with Coulomb interaction strength parameters a) $\tilde c = 0.007$, b) $\tilde c= 0.07$, 
c) $\tilde c = 0.7$ without three-body interaction. The dotted blue line represents the dimer, the solid red lines are the trimers and 
the dashed green lines the tetramer states. The scales on the $x$- and $y$-axes are the same for all plots to make them 
more easily comparable.}
\label{generalplots}
\end{figure}

The case of weak Coulomb interaction is shown in Fig.~\ref{generalplots}~a).
Near the continuum threshold, the dimer shows slight deviations from the universal $1/a^2$ 
behavior. Since we calculated the identical boson case, the scaling factor 
is expected to be $\lambda = 22.7$.
Due to this high scaling factor, we could only resolve the lowest 
two trimers in our calculation.

Comparing the scaling factors of the no-Coulomb case and the weak-Coulomb case, 
however, reveals already some drastic change. First, we note that the scaling
factors calculated at the three-body threshold and in the unitary limit
are the same in the universal limit. In our calculations, however, there
are some non-universal effects due to the fact that we calculate
the ground state without a repulsive three-body force
tuned to push the ground state into the universal region.
We estimate these effects by
calculating the scaling factors in both limits. In the 
no-Coulomb case  with $c_c=0$
the scaling factor is $18 \pm 1$ when comparing the intersections
of the states with the three-body threshold ($a^-$) 
and $23\pm1$ when comparing the square roots of the binding energies
in the unitary limit, where only the numerical errors are quoted.
For the $c_c/c_s = 0.007$ case calculated with the same Hamiltonian,
the corresponding scaling factors become
$55 \pm 5$ and  $32 \pm 3$, respectively.
As discussed above, some slight asymmetry (as in the $c_c=0$ case) is 
expected since the ground state for this Hamiltonian is not fully universal,
but the large discrepancy seen in the weak-Coulomb case cannot be explained
by non-universal behavior of the ground state alone. So, even though the
Efimov plot does not look very different for the two cases at first sight,
a significant breaking of the discrete scaling symmetry
can already be observed. 

Comparing the ratio between binding energy at unitarity $B_3$ and
inverse scattering length at the three-body threshold  $a^-$ is also 
instructive. The most precise universal value from the  literature is
$a^-\sqrt{mB_3} /\hbar = −1.5077(1)$~\cite{deltuva2012recombination}.
Without Coulomb interaction, we obtain 2.1(1) for the trimer ground state and 
1.6(1) for the excited state.
For the case of weak Coulomb interaction ($c_c/c_s = 0.007$), 
we have 2.3(1) for the ground state and 4(1) for the excited trimer.
This shows that the strong deviation we observe for the scaling factor
mainly stems from the  deformation of the excited state by the Coulomb
interaction while the more compact ground state remains relatively unaffected.
This is in agreement with the simple scale argument from Sec.~\ref{sec:intro}.
For stronger Coulomb interaction, however, the ground state
trimer also starts to become deformed and we obtain the factors
4 for $\tilde c = 0.07$ and 40
for $\tilde c = 0.7$.

If the Coulomb interaction is made stronger by tuning up the strength
parameter $\tilde c$, 
the generalised Efimov plot changes significantly as seen in panels
b) and c) of Fig.~\ref{generalplots}.
The excited states move towards positive scattering length
and larger binding energies. They eventually vanish from the spectrum
as can be seen in Fig.~\ref{generalplots} b) and c). 
When considering the fact that near the threshold, states typically
are more dilute this behavior is to be expected. The larger the average
distance between particles, the more  influence the long-range Coulomb
force gains over the short-range attraction. Therefore, one
expects the Coulomb interaction to have the strongest effect for the
shallow states.

Since there is no excited state anymore for $\tilde c=0.07$ in the unitary
limit, the scaling factor between two consecutive trimers is not
well-defined anymore. We can, however, investigate
the evolution of the scaling factor that connects the tetramer to the trimer. 
In the case without Coulomb interaction, we obtain $2.3 \pm 0.1$ for the scaling of the
intersection of the ground state of the tetramer and the ground state of the
trimer with the $N$-body threshold and
$2.4 \pm 0.1$ for the scaling of the square root of the binding energies
for the same states at unitarity. 
This is reasonably close to the
universal value $B_4^{0}/B_3 =4.610(1)$ from Eq.~(\ref{eq:scale-tet}). 
Our result without Coulomb interaction is $5.8 \pm 0.5$. The difference
from the universal value again gives an estimate of the size of
non-universal effects in our calculation.
For the ratio of the scattering lengths corresponding to the
intersections of the states with the $N$-body threshold,
Deltuva obtained $0.4254(2)$ \cite{deltuva2012recombination}.
This should be compared to the inverse of our number, i.e. $0.43 \pm 0.01$.
This agrees much better, because it is extracted from
threshold properties, where non-universal effects are less severe. 

For weak Coulomb interaction ($\tilde c=0.007$), the scaling factors become slightly smaller, 
$2.2 \pm 0.1$ for the intersection with the $N$-body threshold and
$2.3 \pm 0.1$ for the square root of the energies. 
But compared to the large modification of the scaling factor between the ground and excited state of the trimer, the change is negligible. This might be due to the fact that here we 
compare two ground states that have similar distances between the particles
(see Sec.~\ref{sec:structure}) and are therefore affected in similar ways by
the Coulomb interaction.
If we go to $\tilde c=0.07$, the scaling factor becomes even smaller, $2 \pm 0.1$ and $2.1 \pm 0.1$, 
respectively. A caveat is in order for the second of these values, because the ground state 
of the tetramer is not strictly inside the range of our theory at unitarity. It is still 
evident, however, that although the generalised Efimov plot shown in Fig.~\ref{generalplots} c) does 
not much resemble the standard Efimov plot any more, the scaling between the
ground state tetramer and trimer remains qualitatively the same.

In the case of a strong Coulomb interaction, illustrated in Fig.~\ref{generalplots} c), the excited 
states have vanished completely from the region where $E < E_s$. The scaling factor between 
the trimer and tetramer ground states is $1.7 \pm 0.1$ if read off from the
intersections of the states with the $N$-body
threshold. If one were to compare the energies as well, the resulting factor would be 
$1.9 \pm 0.1$, but the tetramer energy is quite far outside the range of $E/E_s < 1$, so 
it should be disregarded. However, just from the first value we can observe that the 
scaling factor continues to decrease, but is still not as drastically influenced as other 
aspects of the Efimov plot.

\section{Structure}
\label{sec:structure}

To better understand the effect of the Coulomb interaction on the 
Efimov spectrum, we analyse the structure of the states as well. 
There are two parts to the analysis. In the first part, we investigate 
the root mean square (rms)
distance between two particles,
$\sqrt{\expval{r^2_{ij}}}$. 
Since we have identical particles and thus no way to single out 
a specific particle pair, this value will be averaged over all 
particle pairs. This will give us an idea about the general size of the 
bound state. 
In the second part we will then supplement this analysis with 2D visualisations 
of the particle density distribution, in order to capture different 
topologies. 

\subsection{Rms Distance Analysis}

\begin{figure}[tb]
\bigskip
\includegraphics{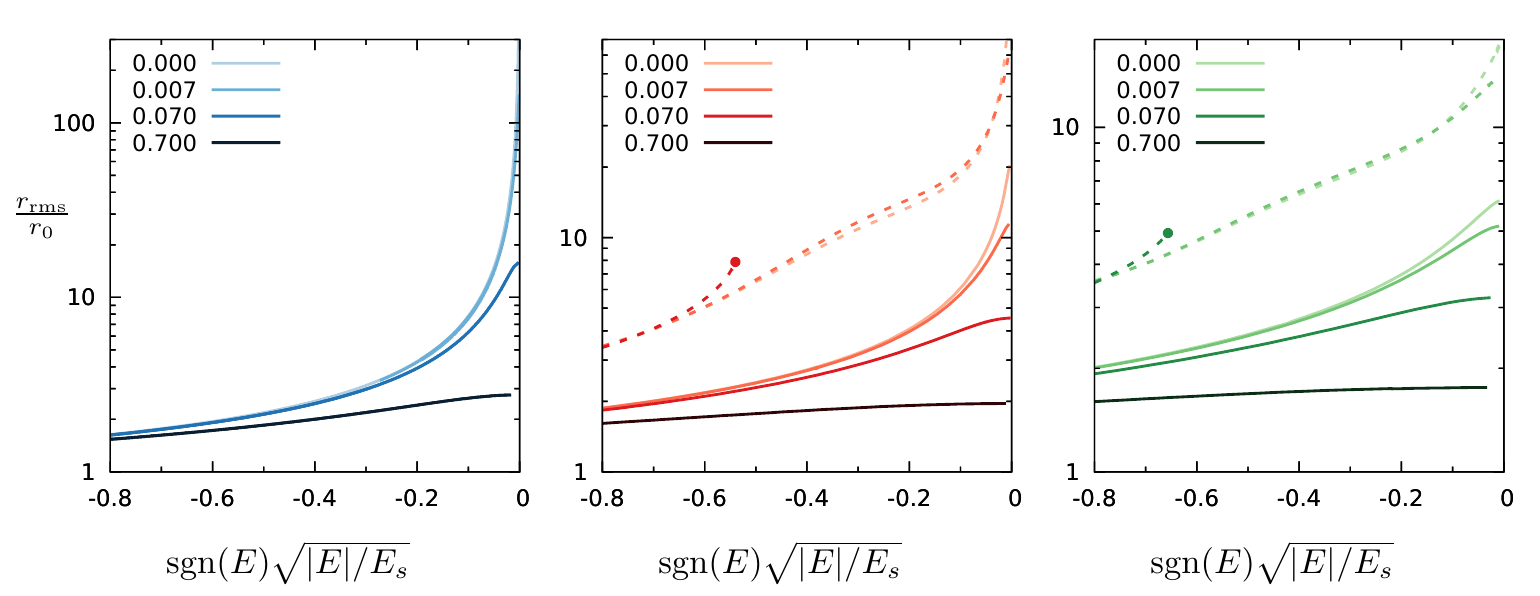}
\caption{Value of the root mean square radius $r_\text{rms}$ of the dimer (left panel), trimer (middle panel)
 and tetramer (right panel) in natural units against  the square root of the
 binding energy in natural units for different strengths of the Coulomb potential $\tilde c$. The value 
 of $\tilde c$ is indicated in the legend. The solid lines 
 represent the ground state and the dashed lines the excited state. There is no excited state for the 
 strong-Coulomb case in the shown region.}
 \label{rmsvsE}
\end{figure}

To calculate the rms pair distance we calculated the matrix elements for $(\sum_{i,j}r^2_{ij})/N_\text{pairs}$
and multiplied this with the wavefunctions for the respective states that we obtained from our calculations.
Then we took the square root of the result to obtain $\sqrt{\expval{r^2_{ij}}}$. 

The results for the dimer are shown in the left panel of Fig.~\ref{rmsvsE} for different strengths of 
the Coulomb interaction. An interesting observation that can be made is that while 
for the no-Coulomb case the dimer seems to become arbitrarily large for smaller 
and smaller binding energies, the presence of the Coulomb interaction leads to 
a plateau in the maximal size. For larger binding energies, the dimer seems almost 
unaffected even in the strong-Coulomb case. This explains the feature observed in the 
generalised Efimov plots, where the dimer follows almost a straight line which corresponds 
to the universal dimer $\sim 1/a^2$ until it becomes too shallow and curves up to the 
break-up threshold very steeply. There seems to be a maximal separation distance up to 
which the short-range potential can support a bound state against the Coulomb repulsion. 
When this maximal separation is reached, the state vanishes into the threshold very quickly. 

This behavior is similar for the first excited trimer (middle panel of Fig.~\ref{rmsvsE}) 
which becomes very large close to threshold 
without the Coulomb interaction, but reaches its maximal size much more quickly with 
even a very weak Coulomb interaction present. For even stronger Coulomb interaction, the excited state vanishes into the 
dimer+atom threshold, so it is not possible to further trace the behavior. 
 Shortly before vanishing, 
the rms radius becomes larger, which is probably due to one of the atoms separating 
from the other two.
The ground state stays generally much smaller than 
the excited state, because it is further away from unitarity, i.e. at smaller negative values of the 
scattering length, for the same binding energy.
This qualitative behavior is in agreement with the results of Fedorov et al.
for three-body halos with Coulomb~\cite{Fedorov:1994}. 

An often used criterion for Efimov states is that the size of the state $R$ 
should be much larger than the range of the interaction
(cf.~\cite{deltuva2010}).
If we take $r_0$ as the range of the short-range interaction, we can
see that the Coulomb barrier forces the states to become smaller with
increasing Coulomb interaction strength. This pushes 
the states out of the universal region and away from the Efimov criterion
$r_\text{rms} \gg r_0$.  
For weak Coulomb interaction the criterion is still well satisfied,
but for stronger Coulomb repulsion the rms radius is of the order of $r_0$,
which also explains the large modifications of the spectrum shown in
the previous section. 

For the tetramer, a similar picture emerges, as is evidenced in the right panel of Fig~\ref{rmsvsE}. 
The rms distance of the ground state again
becomes smaller when the Coulomb interaction becomes stronger. The excited tetramer is 
larger than the  ground states for all strengths of the Coulomb interaction, but vanishes 
into the trimer+atom threshold if the Coulomb interaction becomes too strong.

\subsection{Contour Plot Analysis}

In this section, we show the particle distribution of the trimers as a 2D heatmap. 
The sampled particles are first aligned such that the principal 
axis with the smallest moment of intertia lies on the $y$ axis.
The configurations are then mirrored 
along the $x$ and $y$ axis such that there is always one particle in the right upper quadrant 
and two particles in the lower half of the plot. This procedure helps to visualise the structure 
of the sampled state, because it sorts the samples such that for identical particles the two particles closest to each other 
are always in the lower two quadrants with the third particle in the upper right quadrant. The heatmap allows to analyse 
the structure of the state without being disturbed by the different possible orientations of the system in space.

\begin{figure}[tb]
\includegraphics{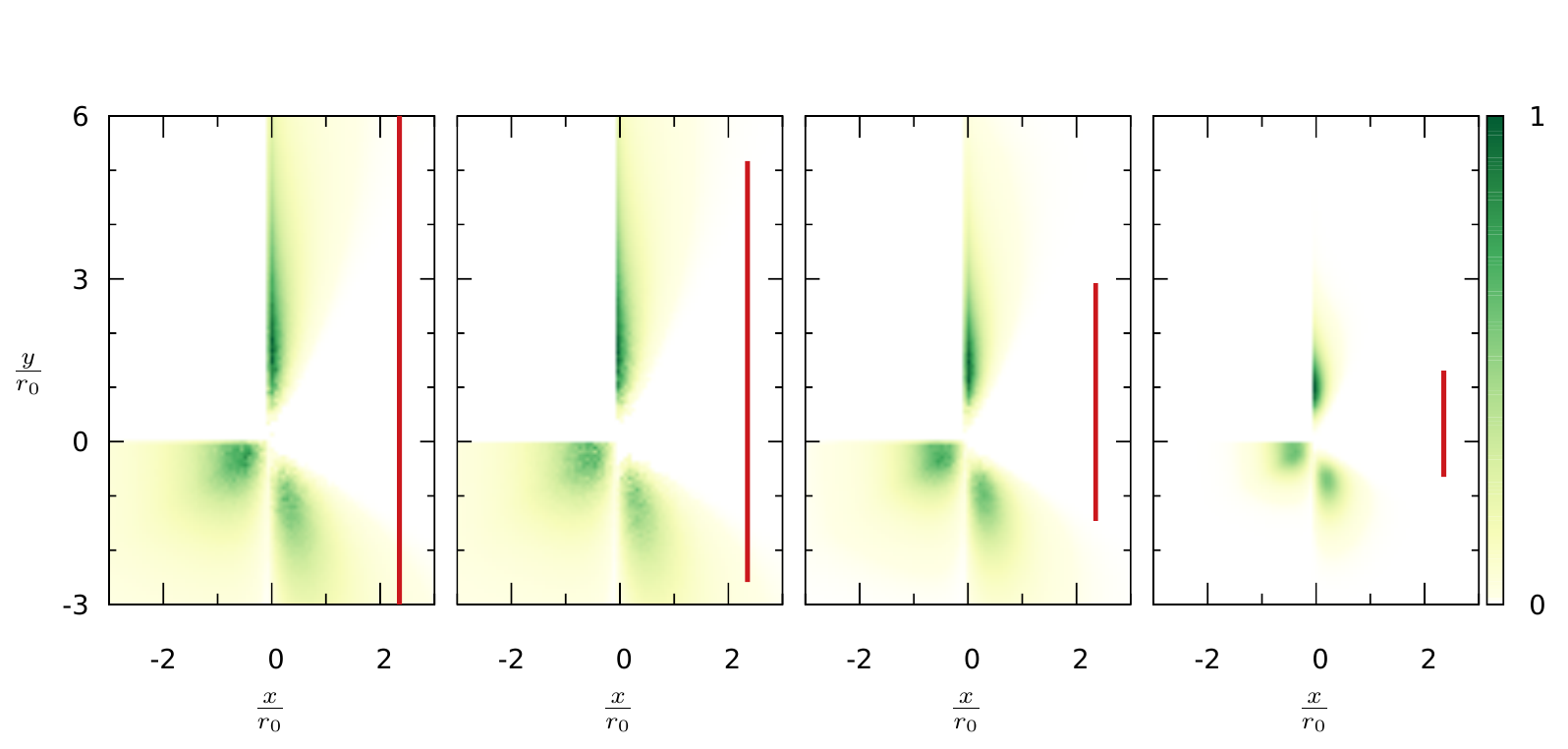}
 \caption{Contour plot of the ground state for different $\tilde c$, at the point where 
 the binding energy is $0.05E_s$. From left to right $\tilde c = 0, 0.007, 0.07, 0.7$. 
 The red bar shows $r_\text{rms}/r_0$ for the corresponding state. The $x$ and $y$ axes show 
 distance in units of $r_0$. The plot shows sampled 
 distributions of particles with their center of mass in the origin in the principal axis frame.
 They were rotated such that the principal axis with the smallest moment of inertia is on the 
 $y$ axis, and if necessary mirrored on the $x$ or $y$ axis to ensure that the left upper quadrant is empty. 
 The colors correspond to the summed probabilities of finding a particle in any given bin after this 
 procedure with white being zero probability and dark green being the highest probabilty in a given plot.  }
 \label{contourground}
\end{figure}

In Fig.~\ref{contourground}, we show the contour plots of the ground state trimer for 
different strengths of the Coulomb potential at a binding energy of $0.05E_s$,
which was chosen because it represents a rather shallow state. An interesting observation 
that can be made from the plots is that for increasing Coulomb interaction strength the particle clouds 
become less fanned out, while the positions with the highest probability density stay 
almost the same. So the differences in rms radii stem in a large part from the removal 
of dilute but comparatively improbable configurations and less from a reduction in size 
of the highest-probability configuration.

\begin{figure}[tb]
 \bigskip
 \includegraphics{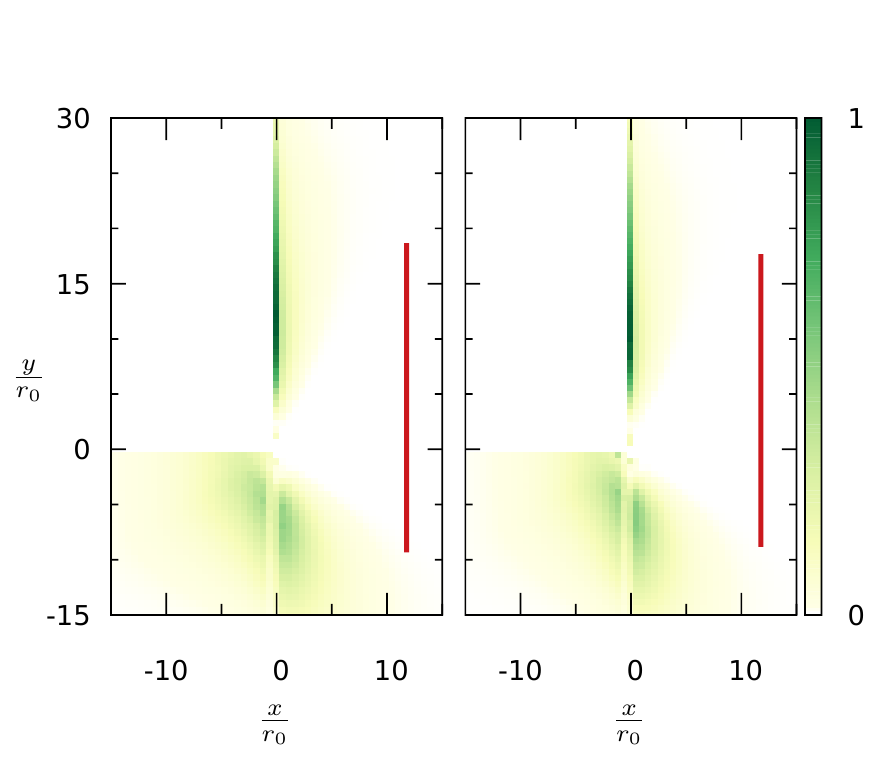}
 \caption{Excited states without Coulomb interaction (left) and with weak Coulomb interaction ($\tilde c = 0.007$)
 (right), at the point where 
 the binding energy is $0.05E_s$. For more information on the type of plot, refer to Fig.~\ref{contourground}
 }
 \label{contourex}
\end{figure}

The excited state follows the same trend as can be seen in Fig.~\ref{contourex}, but since 
the excited state does not exist for stronger Coulomb forces, it is less clear. As has 
already been shown to be the case without the Coulomb interaction \cite{Naidon:2016dpf}, for weak Coulomb interaction the 
shape of the excited state also resembles an elongated triangle. 
Unfortunately, with the parameters used here the excited state vanishes quite quickly 
and therefore its behavior cannot be traced towards stronger Coulomb interaction, 
where it might be possible to link its structure with the structure 
of the Hoyle state, which also has been found to resemble an elongated triangle \cite{fedorovjensen1996}.
Tracking the states as they become resonances might make it possible to 
link the states found here with the Hoyle state, which lies above the $4\alpha$ breakup 
threshold.
This is a question that could be addressed in a future project.

\section{Realisation in Systems of \texorpdfstring{$\alpha$}{alpha} Particles}
\label{sec:applications}

As mentioned in section \ref{sec:interaction}, when comparing our results 
to physical systems, we have to fix $r_0$ in some way. Because the charges 
are fixed by nature, fixing $r_0$ gives us the $\tilde c$ universality class 
of the system via $\tilde c =  Z_i Z_j mr_0 \alpha \frac{\hbar c}{\hbar^2} $.

One obvious way to determine $r_0$ is to choose a value for $r_0$ such that 
the Coulomb-modified effective range ($r_\text{eff}^C$)
is reproduced. 
This ansatz has already been used in \cite{schmickler2018} to
describe $^{17}$F$(\frac{1}{2}^+)$ as a dimer of $^{16}$O and a proton,
leading to consistent results for the dimer system. 
To investigate the applicability of our results in bigger nuclear systems,
we will use systems of three and four $\alpha$ particles as an example.
When treating clusters of $\alpha$ particles, our present approach is
of limited use because most clusters of 
$\alpha$ particles are resonances and not bound states with respect
to break-up into $\alpha$ particles. This is, e.g., true 
for $^8$Be and the Hoyle state in $^{12}$C. 
However, there is a  $0^+$ state of $^{16}$O
at $14.032\,\text{MeV}$~\cite{tilleyweller1993} 
(commonly referred to as $^{16}$O ($0^+_5$)), 
approximately $0.4\,\text{MeV}$ below the four-$\alpha$ 
threshold,
that could  be described as an $\alpha$ cluster state. 
In order to investigate whether this state can be 
interpreted as the remnant of a universal tetramer connected to an
Efimov state, we calculate the generalised 
Efimov plot for the three- and four-$\alpha$ system. 

\begin{figure}[tb]
 \bigskip
 \includegraphics{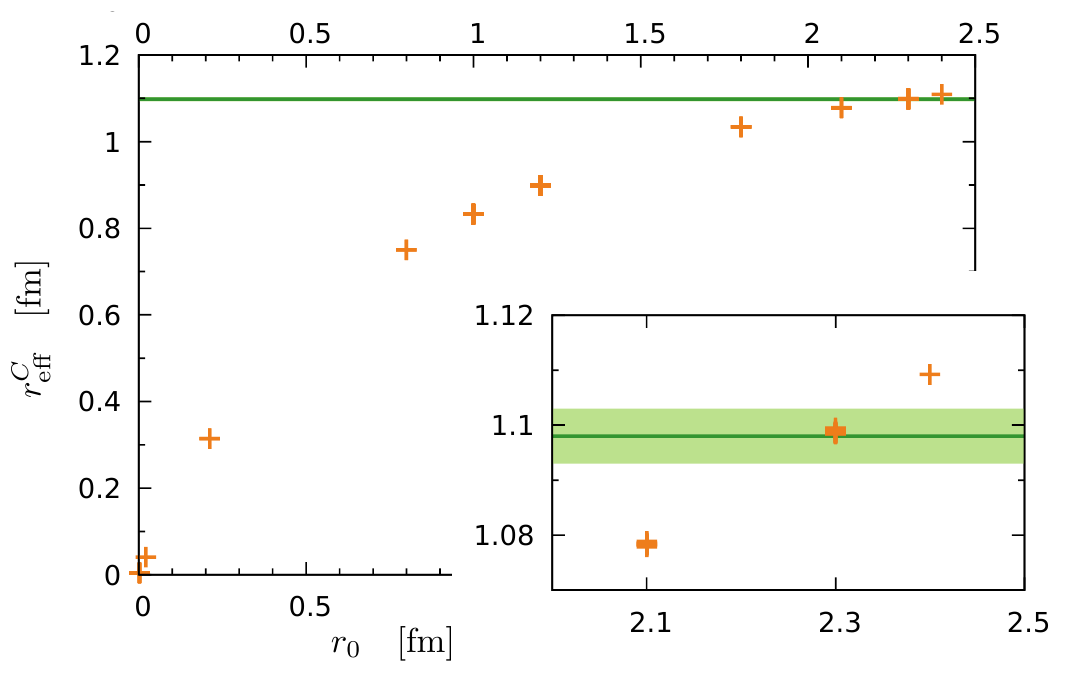}
 \caption{Determination of the effective range at the point of the $\alpha\alpha$ scattering length for different $r_0$. 
 The horizontal line shows the value of the Coulomb-modified effective range in the $\alpha\alpha$ system.}
 \label{reffvsr0alpha}
\end{figure}

As for the $^{16}$O+p system treated in \cite{schmickler2018},
we first determine the $r_0$ that 
reproduces the correct Coulomb modified effective range for
the $\alpha\alpha$ system. This is necessary because 
the Coulomb interaction introduces an additional scale which means that
results for different ranges of 
the Gauss potential $r_0$ are different. We choose
$r_0$ such that $r_\text{eff}^C$ has the correct physical value.
The connection between $r_0$ and $r_\text{eff}^C$
for the  $\alpha\alpha$ system is shown in
Fig.~\ref{reffvsr0alpha}. 
The values for the scattering length and the effective range are 
taken from \cite{higahammer2008}. For Fig.~\ref{reffvsr0alpha},
we select points where the 
scattering length is within the specified errors from \cite{higahammer2008}:
\begin{equation}
  a_C \approx (-1920 \pm 90)\,\text{fm}\,.
\end{equation} 
This leads to the spread of effective ranges $r_\text{eff}^C$
in Fig.~\ref{reffvsr0alpha}. 
The effective range band drawn in the insert of 
Fig.~\ref{reffvsr0alpha}
gives the value from \cite{higahammer2008}, 
$(1.098 \pm 0.005)\,\text{fm}$, including errors.
Thus choosing $r_0$ to be $2.3\,\text{fm}$ reproduces the physical values
of $r_\text{eff}^C$ and $a_C$. 
With this value of $r_0$, $\tilde c \approx 1.27$.

\begin{figure}[htb]
\includegraphics{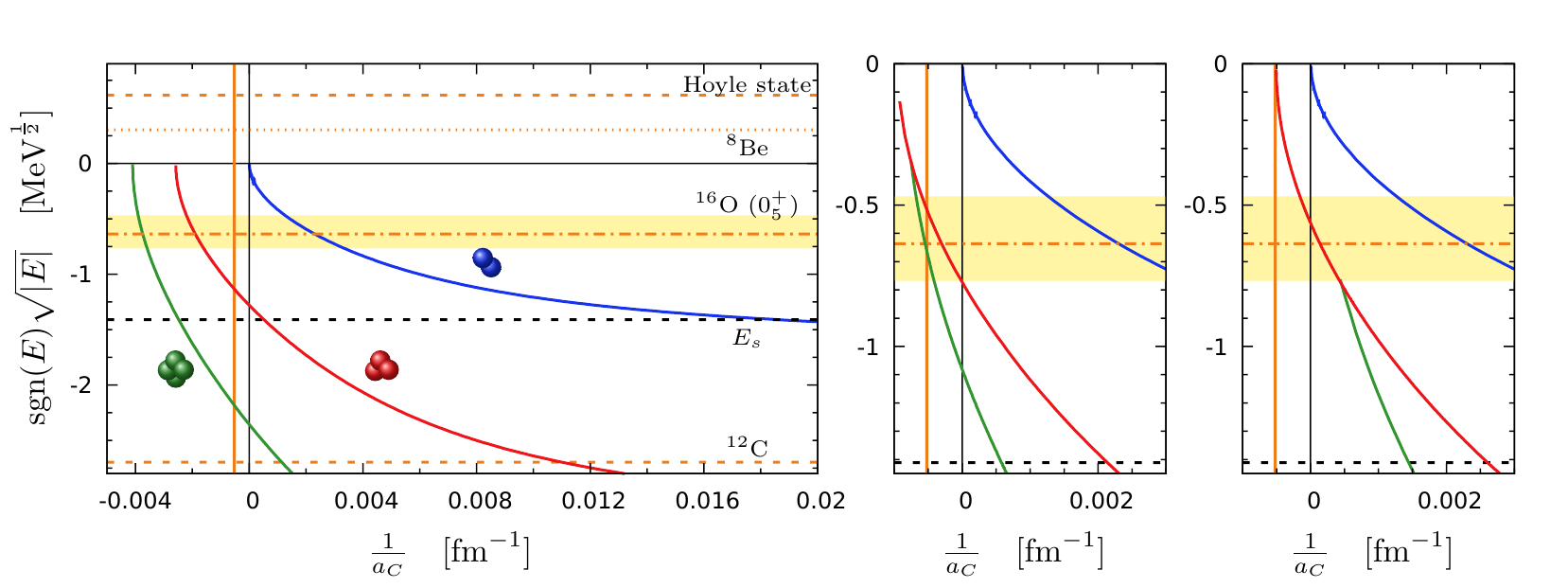}
\caption{\textit{Left panel:} $N\alpha$ system with binding energies and widths of $0^+$ states from Refs.~\cite{tilleyweller1993,kelleypurcell2017,tilleykelley2004}. The dotted horizontal line represents the $^8$Be ground state, 
 the dashed horizontal lines are the ground state of $^{12}$C and the Hoyle state from bottom to top, and the dash-dotted line is
 the $^{16}$O ($0_5^+$) 
 state.
 The vertical solid line is at the 
 $\alpha\alpha$ scattering length $a_{\alpha\alpha} \approx -1920\,\text{fm}$. In addition results for the dimer (blue solid line), trimer (red solid line) and tetramer (green solid line) without three-body force are shown. The dashed horizontal black line marks the natural energy scale $E_s$, which is a measure for the range of validity of our theory. \textit{Middle panel:} Detail of the left plot with the 
 three-body force fitted to reproduce the shallowest bound tetramer state. \textit{Right panel:} Same as the middle panel with the 
 three-body force fitted such that the trimer is unbound at the physical $\alpha\alpha$ scattering length.}
\label{alphaefimov}
\end{figure}

We now turn to the three- and four-$\alpha$ system. It turns out that
with this value for $r_0$, it is impossible to describe the highest
excited state of $^{16}$O ($0^+_5$)
below the four-$\alpha$ breakup threshold
without at the same time having a three-$\alpha$ bound state above
 the $0^+_5$ state of $^{16}$O.
However, such a state does not exist in nature. This is 
illustrated in Fig.~\ref{alphaefimov} where the energies of
the two-, three-, and four-$\alpha$ systems are given as function of $1/a_C$
for physical $r_\text{eff}^C$. The vertical solid line gives the
physical value for $1/a_C$ in the $\alpha\alpha$ system.
Thus we conclude that the highest $0^+$ excited state of  $^{16}$O
cannot be connected to a remnant of an Efimov state, at least not
with our strategy of fixing $r_0$ and omitting higher-order
interactions. This could point to significant contributions of short-range
non-$\alpha$-cluster configurations to this state. 

Moreover, when analysing the rms distance values, we find that $r_\text{rms}$ for the tetramer without three-body force 
at the physical scattering length (corresponding to the left panel of Fig.~\ref{alphaefimov}) 
is only $3.1\,\text{fm}$, which is only barely larger than $r_0 = 2.3\,\text{fm}$. Adding a three-body force 
increases the rms distance between two $\alpha$ particles insubstantially. For $w_0 = 0.775 E_s$, which is 
the value of the three-body potential strength parameter that reproduces the $^{16}$O ($0^+_5$) binding energy (middle 
panel of Fig.~\ref{alphaefimov}), 
$r_\text{rms} = 3.4\,\text{fm}$. This again
suggests that this state is heavily influenced by short-range effects and 
cannot be accurately described in a universal theory of $\alpha$ clusters. 

From the differences between our universal model for $\alpha$ clusters
and the $0^+$ bound states in $^{12}$C and $^{16}$O,
we can conclude that significant non-universal contributions
play a role for the bound states of three and four clusters.

\section{Summary and Outlook}
\label{sec:outlook}

We have investigated the universal properties of systems
with a short-range attractive interaction (modeled by
Gaussian interactions with a range $r_0$) and a repulsive
long-range Coulomb interaction. Such systems can be realized in
nature in $\alpha$ cluster nuclei and few-body systems of ions.
A particular focus was placed on connecting such systems to
Efimov states and their higher-body analogs in the case without
Coulomb interaction and studying their evolution as a function
of the Coulomb interaction strength.

Our results are summarized in generalised Efimov plots where
the scattering length is replaced by the Coulomb-modified
scattering length.
Our plots can
be used to identify universal states of charged particles. 
In natural units, the spectra are fully
universal, i.e. independent of the Gaussian range $r_0$.
Varying the dimensionless Coulomb strength $\tilde c$, we find 
that although the structure of the Efimov plot remains qualitatively the same,
the discrete scaling symmetry is broken strongly already for
very weak Coulomb interaction.
The long-range Coulomb interaction has a strong effect
close to threshold but leaves the deeper states almost unchanged,
which clearly breaks the scaling symmetry. Moreover, the scaling factors
connecting the three- and four-body ground states are also
only weakly affected by the Coulomb interaction.
A study of the structure of these states reveals
that their constituents become more localised
for stronger Coulomb force, while the highest-probabilty 
configuration is almost unaffected.
This can be interpreted as an effect of the rising Coulomb
barrier~\cite{Fedorov:1994}.

Since the Coulomb interaction introduces an additional scale, the
results for different ranges of the Gaussian potential $r_0$ lead to
different physical scenarios. Thus, we have fixed the range $r_0$
to reproduce the Coulomb-modified effective range $r_\text{eff}^C$
to its physical value. However, other strategies to deal with this
problem are possible and will be investigated
in future work. 

As an example, we have applied our model to three- and four-body
systems of $\alpha$ particles.
With our strategy for fixing $r_0$ we find that the
$0^+$ bound state spectrum of $^{12}$C and $^{16}$O
cannot be described by our universal model, at least
without including higher-oder interactions. From our investigations
of the structure of these states, we conclude that
this is related to the small size of these states
which implies significant non-universal contributions.

This leads to the interesting question of whether
the universality   discussed in this work
could be observed in the resonance sector of $^{12}$C and $^{16}$O,
i.e. above the breakup threshold into $\alpha$ particles. In addition,
the universality may be relevant 
for other systems where the Coulomb interaction is weaker
in comparison to the short-range force such as systems of cold
ions \cite{Mathur:1995,Hattendorf:2016}. 
Work in both directions is in progress.

\begin{acknowledgments}
We thank Artem Volosniev for discussions and
D\"orte Blume for providing details on
the contour plots in Ref.~\cite{kunitskizeller15}.
This work was funded by the Deutsche Forschungsgemeinschaft
(DFG, German Research Foundation) –- Projektnummer 279384907 –-
SFB 1245 and the Federal Ministry of Education and Research (BMBF) under
contract 05P18RDFN1.
\end{acknowledgments}

\appendix

\end{document}